\documentclass[conference]{IEEEtran}

\usepackage{cite}
\usepackage{amsmath,amssymb,amsfonts}
\usepackage{algorithmic}
\usepackage{graphicx}
\usepackage{subcaption}
\usepackage{textcomp}
\usepackage{xcolor}
\usepackage{hyperref}
\usepackage{svg}

\def\BibTeX{{\rm B\kern-.05em{\sc i\kern-.025em b}\kern-.08em
    T\kern-.1667em\lower.7ex\hbox{E}\kern-.125emX}}
\begin{document}
\bstctlcite{IEEEexample:BSTcontrol}
\title{Some Initial Guidelines for Building Reusable Quantum Oracles}


\author{\IEEEauthorblockN{Javier Sanchez-Rivero}
\IEEEauthorblockA{\textit{COMPUTAEX} \\
Cáceres, Spain\\
javier.sanchez@cenits.es}
\and
\IEEEauthorblockN{Daniel Talaván}
\IEEEauthorblockA{\textit{COMPUTAEX}\\
Cáceres, Spain\\
daniel.talavan@cenits.es}
\and
\IEEEauthorblockN{Jose Garcia-Alonso}
\IEEEauthorblockA{\textit{University of Extremadura}\\
Cáceres, Spain\\
jgaralo@unex.es}
\and
\IEEEauthorblockN{Antonio Ruiz-Cortés}
\IEEEauthorblockA{\textit{SCORE Lab, I3US Institute, Universidad de Sevilla}\\
Sevilla, Spain\\
aruiz@us.es}
\and
\IEEEauthorblockN{Juan Manuel Murillo}
\IEEEauthorblockA{\textit{COMPUTAEX and University of Extremadura}\\
Cáceres, Spain\\
juan.murillo@cenits.es}
}

\maketitle

\begin{abstract}
The evolution of quantum hardware is highlighting the need for advances in quantum software engineering that help developers create quantum software with good quality attributes. Specifically, reusability has been traditionally considered an important quality attribute in terms of efficiency of cost and effort. Increasing the reusability of quantum software will help developers create more complex solutions, by reusing simpler components, with better quality attributes, as long as the reused components have also these attributes. This work focuses on the reusability of oracles, a well-known pattern of quantum algorithms that can be used to perform functions used as input by other algorithms. In particular, in this work, we present several guidelines for making reusable quantum oracles. These guidelines include three different levels for oracle reuse: the ideas inspiring the oracle, the function which creates the oracle, and the oracle itself. To demonstrate these guidelines, two different implementations of a range of integers oracle have been built by reusing simpler oracles. The quality of these implementations is evaluated in terms of functionality and quantum circuit depth. Then, we provide an example of documentation following the proposed guidelines for both implementations to foster reuse of the provided oracles. This work aims to be a first point of discussion towards quantum software reusability. Additional work is needed to establish more specific criteria for quantum software reusability. 
\end{abstract}


\section{Introduction}

The development of quantum computers and simulators in the NISQ era has opened the door to the design and building of quantum programs that can now be run and tested. Quantum software is beginning to become a reality and with it, the first forums have begun to appear that address the discipline of Quantum Software Engineering (IEEE QSW, ICSE Q-SE, Q-SET, etc). One of the problems to be addressed by this discipline is that of achieving good quality attributes in quantum programs \cite{Zhao, SodhiQA2021}.

The standard ISO/IEC 25010:2011 - Systems and software Quality Requirements and Evaluation (SQuaRE) \cite{ISO25010} identifies \textit{Reusability} as one of the  of the sub-characteristics of the quality attribute \textit{Maintainability}. It is defined as the \textit{degree to which an asset can be used in more than one system, or in building other assets}. Thus, in terms of efficiency of cost and effort, software is of better quality the more reusable it is. Modularity can contribute to reusability in that the asset to be reused needs not be a complete program but a smaller module of whatever it may be (procedure, function, component, service, etc.).

The oracle has been identified as a pattern for quantum algorithms \cite{Leymann-QuantumAlgorithms}. An oracle can be thought as a black box performing a function that is used as an input by another algorithm \cite{oracles_as_black_boxes}.  Although oracles are not real black boxes as they are known in classical software development, the above means that how an oracle works is not a matter of concern for the algorithm that uses it. Following the principles of Separation of Concerns \cite{Dijkstra1982}, they can be designed in isolation and then applied in solving different problems. These features make oracles a good candidate asset for quantum software reuse. 

However, oracles are not reusable by themselves. Software reuse involves building software that is reusable by design and building with reusable software \cite{ReuseMili}. So, as in classical systems programming, quantum programmers must apply techniques that make the code they build reusable.   Also, practices that promote the quantum code reuse are also necessary. 

Proper documentation is a necessity for the systematic reuse of software \cite{ReusableComponents}. Thus, this work focuses on how to properly document quantum software and more specifically quantum oracles. 
Furthermore, some guidelines for  documenting of oracles are proposed. To demonstrate the use of the guidelines two different implementations of the same oracle are provided. These oracles responds to the \textit{range of integers}, i.e., applied to a state in superposition that encodes integers and given two integers $n_1$ and $n_2$ ($n_1 < n_2$), both oracles mark in phase all those integers greater or equal than $n_1$ and less or equal than $n_2$.


This work is part of a research direction that aims to provide quantum programmers with operations that can be composed at a higher level of abstraction than quantum gates. Thus, for example, one of the implementations of the \textit{range of integers} oracle offered in the paper corresponds to a reuse and composition of two oracles \textit{less than} and \textit{greater than}. We find that for programmers to be able to make proper use of these more abstract operations, their implementations need to be well documented.


The remainder of this paper is structured as follows. Section \ref{sec:background} introduces some background and related works about oracles and their usage. Then, section \ref{sec:RangeOracles} introduces two different implementations for the \textit{range of integers} oracle. One of them is based on the composition of two oracles \textit{less than} and \textit{greater than}. The second one corresponds to a direct implementation optimizing the depth of the produced quantum circuit. Section \ref{sec:guidelines} analyses the documentation needs for oracles to be reused. We analyse three different levels of re-usability said reuse of the ideas inspiring the oracle, reuse of the oracle and reuse of the classical function that builds the implementation of the oracle. Finally, section \ref{sec:conclusions} gives some conclusions and explores future works. 

\section{Background and Related Work}\label{sec:background}

While oracles can be viewed as black boxes, it is important to know how they will affect the quantum state they are applied to. In terms of their effects on quantum states there are two main types of oracles \cite{typesOracles2019}: probability and phase oracles. The former are common in quantum optimization procedures. The latter are used in quantum algorithms and encode a function in the phase of the quantum states. 


A well-known example of a quantum algorithm that uses an oracle is Grover's algorithm \cite{grover1}. This quantum algorithm can search a value in an unordered data sequence faster than any classical algorithm. In order to do that, it needs an oracle that encodes the desired value. There are many other quantum algorithms that use oracles such as Deutsch-Jozsa \cite{Deutsch1992RapidSO}, Simon \cite{Simon_1997} or Bernstein-Vazirani \cite{Bernstein_Vazirani}. 


However, the most famous quantum algorithm is probably Shor's  \cite{shor1994algorithms}.  By using this algorithm, it is possible to find the prime factors of a number exponentially faster than with any known classical method \cite{politi2009shor}. The different operations of this algorithm can be understood as oracles as well \cite{Leymann_2020}. In fact, Shor employs the Quantum Fourier Transform, previously described in \cite{QFTcoppersmith}. This can be seen as one of the first cases of quantum software reuse.

It is desired that reusable software have the best possible quality attributes. This is specially relevant in quantum software as actual quantum computers are prone to decoherence. One of the crucial factors regarding reliability of results is the depth of circuits. As depth is a measure of the execution time of a given circuit, the deeper the circuit, the greater the exposition to noise and lower its reliability \cite{depth-decoherence}. Consequently, disregarding the depth of the circuit may arise results indistinguishable from noise \cite{Preskill2018quantumcomputingin}. Therefore, reducing the depth of oracles improves their quality attributes and their reusability.


In this paper we present an example of quantum software reutilisation by combining existing quantum software to build a range of integers phase-marking oracle. We present two different implementations of the same oracle to showcase the need for a good description in order to have easily reusable oracles. Based on this example, we provide some guidelines for making reusable oracles. 

\section{Oracle for Range of Integers} \label{sec:RangeOracles}


In this section we present two different implementations of what is, in principle, the same oracle. We provide the documentation of both implementations. These two different implementations exemplify how already existing oracles can be reused to achieve a different purpose. 


The aim is to design a phase-marking oracle which, assuming quantum states encode natural numbers (including 0), gives a $\pi$-phase to those numbers within a given range, $[n_1, n_2]$. This oracle ideally would encode the following function:

\begin{equation}\label{func:1}
    f(x) = \left\{
    \begin{array}{cl}
        -1 & \text{ if } x\in [n_1, n_2]  \\
         1 & \text{ otherwise} 
    \end{array}
    \right.
\end{equation}





These two implementations of a phase-marking oracle for a range of integers have been developed by reusing existing quantum software. We have used, in all cases, the linear-depth multi-controlled $Z$-gate as provided in \cite{multicontrol2022}. The less-than oracle presented in \cite{sanchez2023automatic} is also used in both implementations. The quantum addition presented in \cite{draper2000addition} is used in implementation B. As all these quantum software pieces have a depth with polynomial growth with the number of qubits, the resulting compositions of these oracles have polynomial depth. All the coding has been done in Qiskit \cite{Qiskit}.

In the next subsections we will explain how these reused pieces of quantum software are combined to get the range of integers oracle and we will present a comparison on both implementations to showcase the differences between them. The code for both implementations as well as the data of experiments conducted in this work can be found in \hyperref[sec:code-data]{Code and Data}.

\subsection{Implementation A: two less-than oracles}\label{subsec:implementationA}
The first way of implementing a range of integers oracle is by combining two less-than oracles. If the aimed range is $[n_1, n_2]$, notice the close interval, the oracle which marks these states can be obtained by applying oracles less-than $n_2+1$ and less-than $n_1$ (equivalent to a greater-than oracle and a global phase). It is noticeable that this oracle marks the desired states regardless of the order. This happens because the states smaller than $n_1$ are marked twice, so the second phase-marking applied to already marked states return them to $0$-phase.

This implementation does perform the function \ref{func:1} for any input state. The unitary matrix of this oracle is \ref{matrixA} where $-1^\dag$ and $-1^\ddag$ are in the positions $(n_1, n_1)$ and $(n_2, n_2)$, respectively.

\begin{equation}
\begin{pmatrix}\label{matrixA}
1 &  &  &  & &  \\
 & \ddots & &  & &  \\
 &  & -1^\dag & & \multicolumn{3}{c}
 {\raisebox{\dimexpr\normalbaselineskip+.7\ht\strutbox-.5\height}[0pt][0pt]{\scalebox{2}{$0$}}} \\
 &  & & \ddots  &  &  \\
 &  &  & & -1^\ddag & &   \\
 &  &  & && \ddots & \\
 \multicolumn{3}{c}{\raisebox{\dimexpr\normalbaselineskip+.7\ht\strutbox-.5\height}[0pt][0pt]{\scalebox{2}{0}}} &  & & & 1  \\
\end{pmatrix} 
\end{equation}

\begin{figure*}[h] 
\begin{subfigure}{0.25\textwidth}
\includegraphics[height=8.5cm]{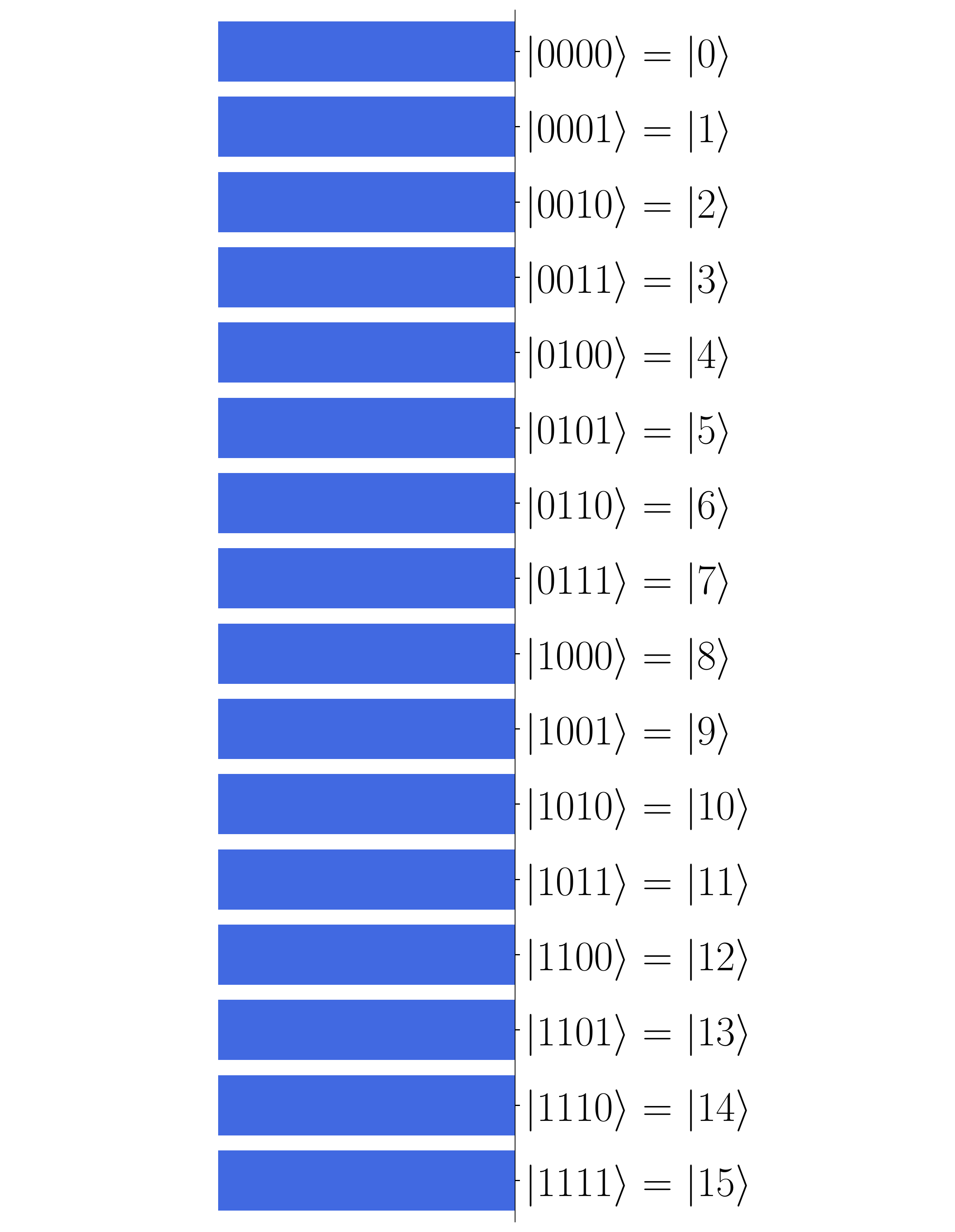}
\end{subfigure}\hspace*{\fill}
\begin{subfigure}{0.35\textwidth}
\includegraphics[height=8.5cm]{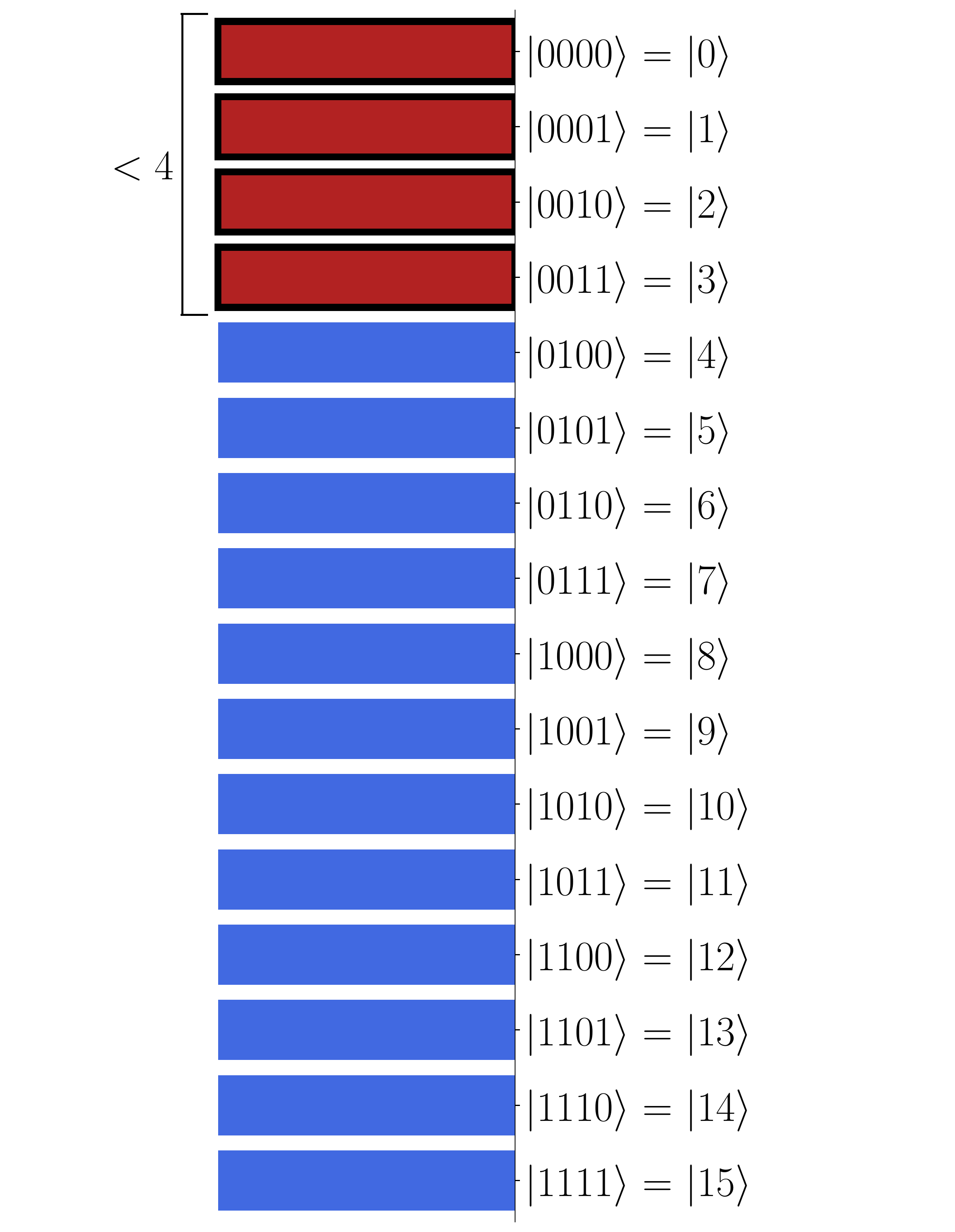}
\end{subfigure}
\begin{subfigure}{0.32\textwidth}
\includegraphics[height=8.5cm]{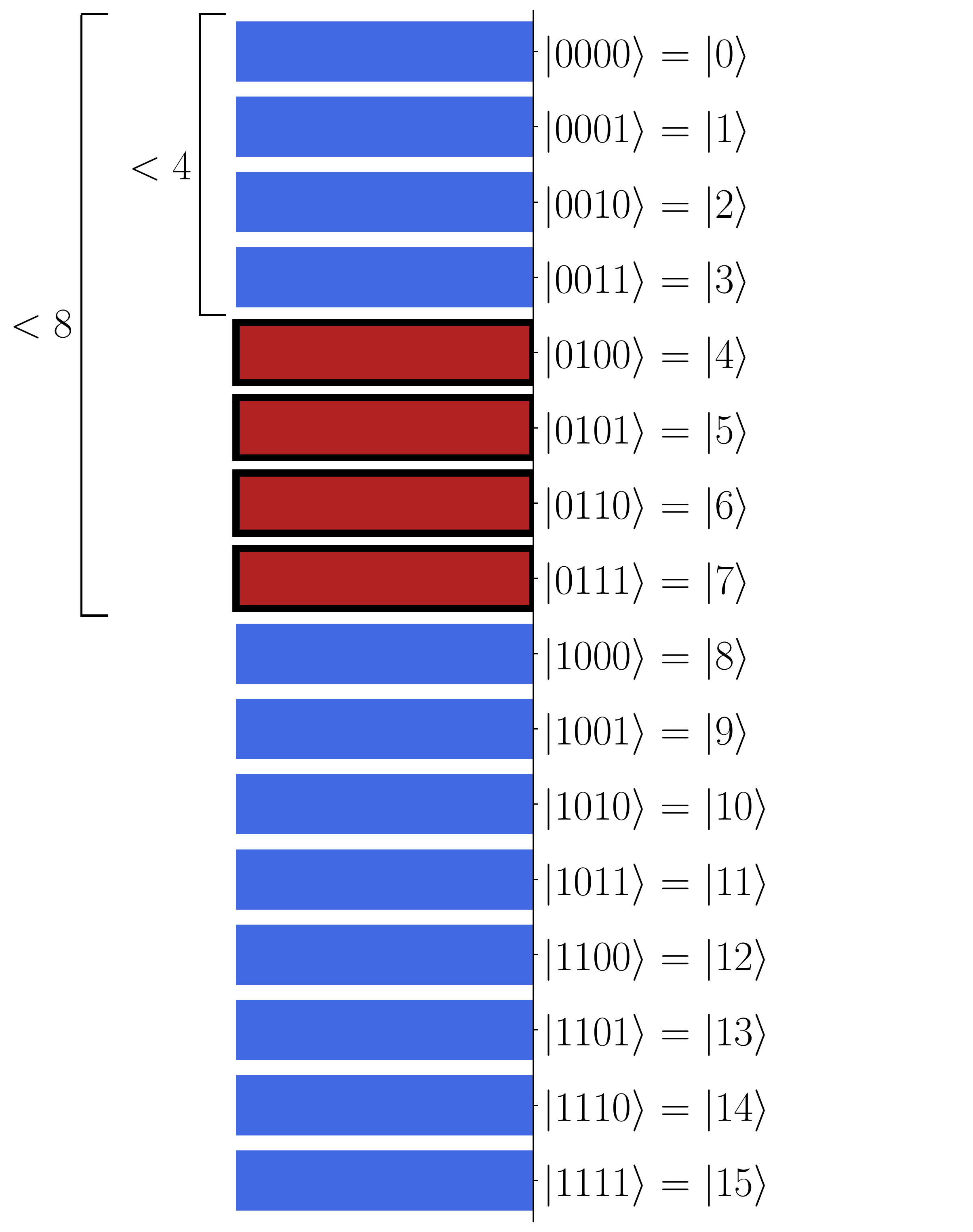}
\end{subfigure}

\medskip
\begin{subfigure}{0.25\textwidth}
\centering
\includegraphics[height=4cm]{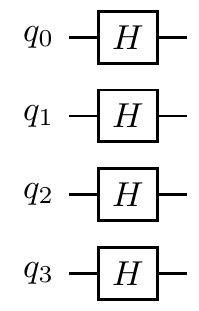}
\caption{Full superposed state and circuit for full superposition.} 
\label{subfig:superpositionA}
\end{subfigure}\hspace*{\fill}
\begin{subfigure}{0.35\textwidth}
\centering
\includegraphics[height=4cm]{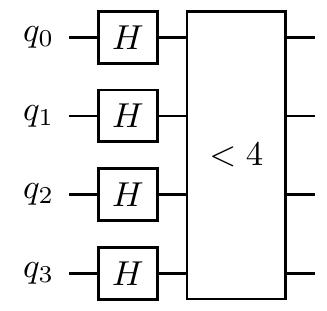}
\caption{States smaller than $4$ marked and circuit after applying less-than 4 oracle.} 
\label{subfig:less-thanA}
\end{subfigure}
\begin{subfigure}{0.32\textwidth}
\centering
\includegraphics[height=4cm]{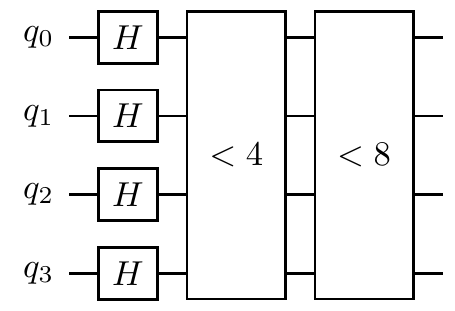}
\caption{States within range $[4, 7]$ and circuit after applying less-than 4 and less-than 8 oracle.}
\label{subfig:rangeA}
\end{subfigure}

\caption{Implementation A of range of integers oracle for $[4, 7]$.} \label{fig:implementationA}
\end{figure*}

An example of this oracle is shown in Figure \ref{fig:implementationA}. The upper part of the figure corresponds to the quantum states after applying the circuits displayed in the bottom part. In blue with no border, the states with a $0$-phase, in red with thick border, the states with a $\pi$-phase. Subfigure \ref{subfig:superpositionA} represents the states after the initialisation to full superposition and the corresponding circuit. Subfigure \ref{subfig:less-thanA} shows in red with thick border the quantum states marked after applying a less-than $4$ oracle to the circuit. Subfigure \ref{subfig:rangeA} shows the states within the desired range successfully marked after a less-than $8$ oracle is applied to the circuit.

\subsection{Implementation B: less-than oracle and displacement}\label{subsec:implementationB}
The second way of implementing this oracle is by applying a less-than oracle followed by a displacement of the marked states by using quantum addition. If the aimed range is $[n_1, n_2]$, the oracle can be obtained by applying the oracle less-than $n_2-n_1+1$ followed by quantum addition of $n_1$. Contrary to the first implementation, the order of the oracles in this implementation must always be as explained.

However, this implementation possesses some drawbacks. A hard condition on the input state is needed as the oracle only performs function (\ref{func:1}) given a full superposed input state without relative phases, formally:
\begin{equation}
    \dfrac{1}{\sqrt{N}}\sum_{i=0}^{N-1} |i\rangle
\end{equation}
where $N = 2^n$ being $n$ the number of qubits. The unitary matrix of this implementation is matrix \ref{matrixB} where $-1^\dag$ and $-1^\ddag$ are in the positions $(n_1, 1)$ and $(n_2, n_2-n_1+1)$, respectively.

\begin{equation}
\begin{pmatrix}\label{matrixB}
 0  &  & \cdots &  &  0 & 1 & & \\
 \vdots & \ddots & & & & \ddots & \ddots &  \\
 0  &  & 0 & & & & \ddots & 1\\
 -1^\dag & \ddots & \vdots & \ddots & \multicolumn{2}{c}{\raisebox{\dimexpr\normalbaselineskip+.7\ht\strutbox-.5\height}[0pt][0pt]{\scalebox{2}{0}}}  & &  0 \\
 & \ddots  & 0 & & \ddots & & &  \\
 &  & -1^\ddag & \ddots &  & 0 & & \vdots\\
 & &  & \ddots & 0 & & \ddots &\\
 \multicolumn{3}{c}{\raisebox{\dimexpr\normalbaselineskip+.7\ht\strutbox-.5\height}[0pt][0pt]{\scalebox{2}{0}}} & & 1 & 0 & \cdots & 0 \\
\end{pmatrix} 
\end{equation}

\begin{figure*}[h] 
\begin{subfigure}{0.25\textwidth}
\centering
\includegraphics[height=8.5cm]{images/full_superposition_state.pdf}
\end{subfigure}\hspace*{\fill}
\begin{subfigure}{0.35\textwidth}
\includegraphics[height=8.5cm]{images/less_than_4.pdf}
\end{subfigure}
\begin{subfigure}{0.32\textwidth}
\includegraphics[height=8.5cm]{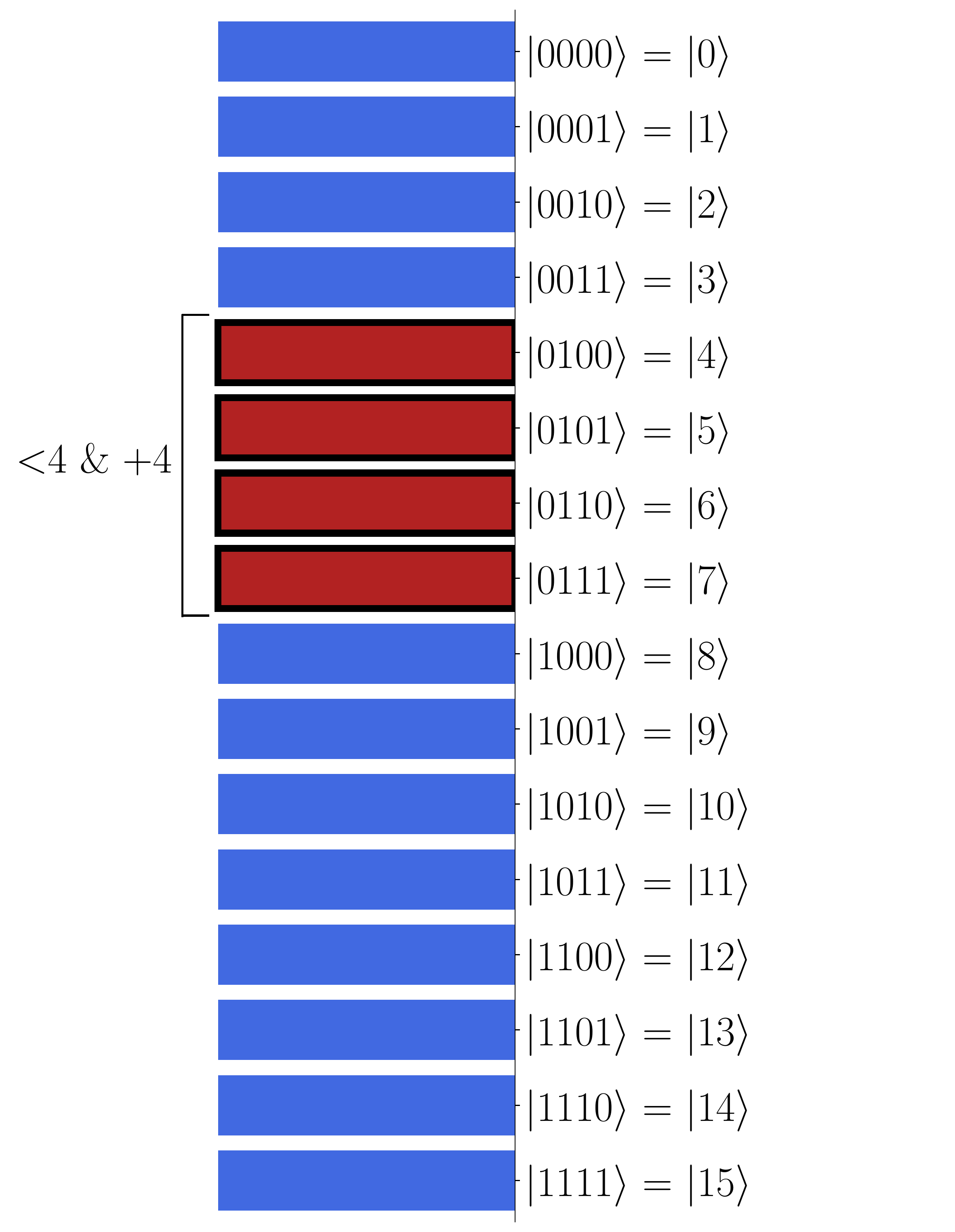}
\end{subfigure}

\medskip
\begin{subfigure}{0.25\textwidth}
\centering
\includegraphics[height=4cm]{images/superposition.pdf}
\caption{Full superposed state and circuit for full superposition.} 
\label{subfig:superpositionB}
\end{subfigure}\hspace*{\fill}
\begin{subfigure}{0.35\textwidth}
\centering
\includegraphics[height=4cm]{images/less-than-4.pdf}
\caption{States smaller than $4$ marked and circuit after applying less-than 4 oracle.} 
\label{subfig:less-thanB}
\end{subfigure}
\begin{subfigure}{0.32\textwidth}
\centering
\includegraphics[height=4cm]{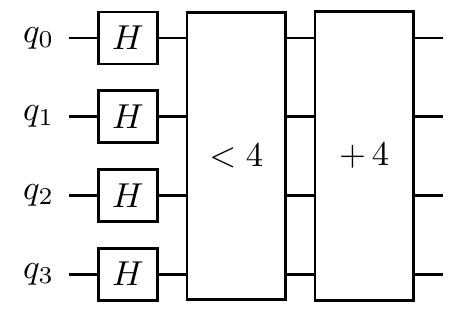}
\caption{States within range $[4, 7]$ and circuit after applying less-than 4 oracle and addition of 4.}
\label{subfig:rangeB}
\end{subfigure}

\caption{Implementation B of range of integers oracle for $[4, 7]$.}\label{fig:implementationB}
\end{figure*}

An example of this oracle is shown in Figure \ref{fig:implementationB}. The upper part of the figure corresponds to the quantum states after applying the circuits displayed in the bottom part. In blue with no border, the states with a $0$-phase, in red with thick border, the states with a $\pi$-phase. Subfigure \ref{subfig:superpositionB} represents the states after the initialisation to full superposition and the corresponding circuit. Subfigure \ref{subfig:less-thanB} shows in red with thick border the quantum states marked after applying a less-than $4$ oracle to the circuit. Subfigure \ref{subfig:rangeB} shows the states within the desired range successfully marked after a less-than $4$ oracle and addition of 4 are applied to the circuit. This addition may be seen as a displacement of the already marked states.

\subsection{Depth comparison of both implementations}
To properly compare the depth of these two methods, we have transpiled both circuits to the gate set used in one of the IBM quantum computers (\textit{ibm\_washington}), using the fake provider \footnote{Fake providers are built to mimic IBM quantum systems, they have the same properties (gate set, coupling map, etc.) as the real devices.}.  

\begin{figure*}[h]
    \centering
    \includegraphics[width=0.9\linewidth]{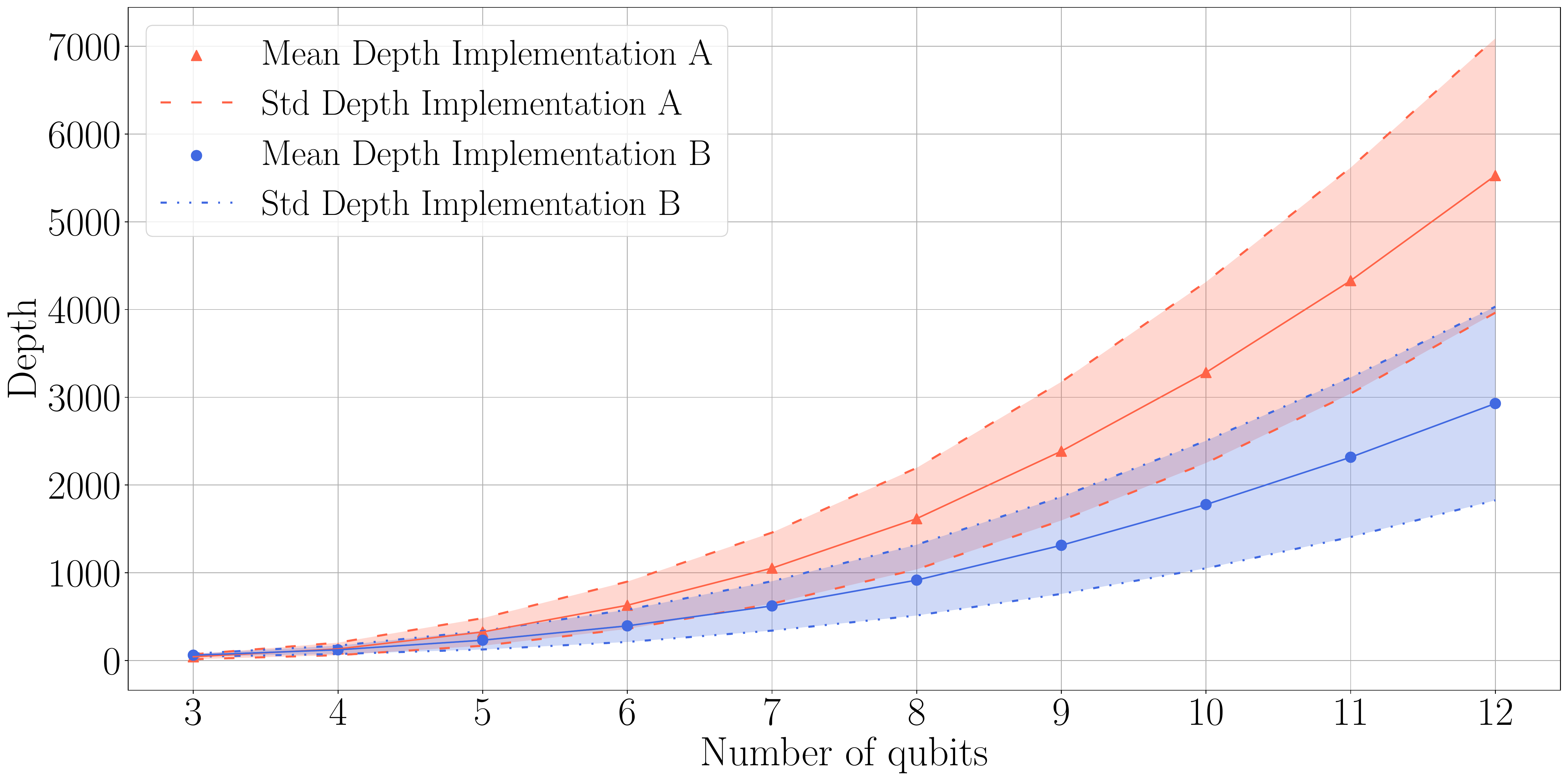}
    \caption{Depth comparison of implementations A and B.}\label{fig:depthcomparison}
\end{figure*}

In order to do the comparison we generate a range of integers circuit for each possible interval $[n_1, n_2]$ where $0<n_1<n_2<N-1$, where $N$ is the total number of states. We have conducted this analysis from 3 to 12 qubits, both included. Figure \ref{fig:depthcomparison} shows the comparison of the two implementations of the range of integers oracle. While both implementations have polynomial depth on the number of qubits, it can be noted that Implementation B has a lower depth than Implementation A. It is more suitable to be used in NISQ devices given a full superposed initial state without relative phases. However, Implementation A, as stated, does not have this constraint and would work on any input state.

\section{Guidelines for Making Reusable Oracles}\label{sec:guidelines}

From the above examples, it is clear that is not enough to document the function that an oracle is performing in order for them to be reused later. Both oracles perform the same function in the adequate input, however choosing one of them over the other can have a significant impact in the final quantum software. From the current version of the oracles, it seems clear that implementation B of the oracle is the preferred one due to the lesser depth of the circuit. However, this could change if a more efficient implementation of the less than oracle that is used twice to create implementation A is discovered. For developers to be able to effectively reuse quantum oracles, they should be aware of all the aspects of the oracle that could have an impact in their software.

In order to facilitate the creation of such reusable oracles, and based on the experience of the creation of the two implementations of the range of integers oracle, in this section we present some initial guidelines for making reusable oracles. We present these guidelines organized in three subsections that represent three different levels of oracle reuse in quantum software.

First, we propose some guidelines for making the fundamentals ideas inspiring the oracle easier to reuse. Given the current state of quantum software development, oracles are usually based on algorithms with low abstraction levels. As has been demonstrated in classical software reuse, this kind of low abstraction level patterns or algorithmic paradigms, if well understood, can help developers in the development process.

Then, we consider two different levels of specification for the reuse of the oracles. On one side, we consider the classical algorithm that is used to create the oracle circuit. This algorithm is needed for generating the actual quantum circuit that would be used as part of the quantum program, adapted to the required size of the input. And on the other side, we considered the oracle itself. This will be the final piece of software that could be reused to create more complex functions by composing them with others and, therefore, developers should be made aware of how to reuse it, how it can be composed, its effects on the quantum state, the impact in the resulting circuit and many other elements needed for effective reuse.

\subsection{Reuse of the Ideas Inspiring the Oracle}

As is well known from classical software, all approaches to software reuse refers to some form of abstraction \cite{krueger1992software}. One of the higher level abstractions in classical software are algorithmic paradigms, a generic model or framework that underlies the design of a class of algorithms. Oracle, as patterns for quantum algorithms can be considered as an algorithmic paradigm and, therefore, a prime candidate for reuse. 

Nevertheless, a high abstraction level is not enough. Any approach to reusability should help software developers locate, compare, and select reusable software artifacts \cite{krueger1992software}. By treating oracles as black box functions, although as mentioned above they are not, we might provide enough information for quantum software developers to understand how an oracle can be used. However, as demonstrated by the presented examples, is not enough if they need to compare them and select the best suited one. 

If we constrain ourselves to the function of oracles A and B, there is not enough information to determine which oracle is better for a given situation. For this specific example, it can be argued that, since the depth of one of the oracles is always smaller, there is no need to select one of the oracles, as the answer will always be the most optimized one. We believe this to be a shortsighted approach given the current status of quantum devices and software. The current diversity in quantum hardware could make a more efficient oracle to perform worse in a given quantum computer or advances in quantum algorithms or transpilers could make a less efficient oracle suddenly improve its performance over others.

For this reason, the first proposed guideline for oracle reuse is that it is not enough to document the function that oracles perform. The underlying components used to create an oracle need to be detailed if we want the oracle to be reused effectively. In \hyperref[fig:oracledocA]{Documentation of implementation A} and \hyperref[fig:oracledocB]{Documentation of implementation B} the Oracle as a Black Box and Oracle as its components show the difference between the two types of oracle documentation.


Additionally, by providing a more detailed description of the oracle we not only make them more reusable by themselves, we allow developers to reuse the ideas inspiring the oracle for the creation of other oracles. An example of this can be found in \hyperref[subsec:implementationA]{implementation A}. As mentioned above, one of the key ideas of this implementation is that if a given state is marked two times (in this case by both less-than oracles) it returns to $0$-phase. This is not an obvious behavior of the less-than oracle, and by making developers aware of it they could be inspired for the creation of new software that can take advantage of this property. 

Taking this into account we can propose another guideline for creating reusable oracles. To foster reuse, a developer must be able to navigate through, and reason about, the source code and its dependencies in order to identify program elements that are relevant \cite{holmes2013systematizing}. In the specific case of quantum oracles this means that the ideas inspiring them should be thoroughly documented alongside the oracle for the developer to be able to reason about them and get inspired for the creation of new oracles based on such ideas.

Another example of this kind of reuse can be found in \hyperref[subsec:implementationB]{implementation B} of the proposed oracle. In that oracle, we make use of the quantum addition to shift the states the marked states to the desired one. 

\subsection{Reuse of the Algorithm that Creates the Oracle}

Once the ideas inspiring an oracle are clearly documented, another relevant level of reuse to consider is the reuse of the function that creates the oracle circuit. For most oracles, a classical algorithm is used in order to generate a specific quantum circuit that implements the oracle for a given size of the input. As such, to foster reuse its documentation should focus on some of the metrics that traditionally help software reuse like the ratio of input/output parameters or the ratio of comments \cite{frakes1996software}.

Additionally, some aspects have to be considered that are specific of quantum software in general and oracles in particular. Specifically, a relevant guideline when documenting for reuse an algorithm that creates oracles is to differentiate between input parameters that are only used by the algorithm that those that are used by the oracle itself.

One example of such parameters in the case of the range of integers oracle are the two specific integers that define the limits of the range. These values are used by the function to create the quantum circuit. However, once the circuit is generated those are not input parameters of the oracle. The circuit that implements the oracle for the range $[4, 7]$ will only work for those values and a completely different circuit should be created for the range $[3, 8]$ or any other range. 

How this kind of parameters of the function but not of the oracle work may not be obvious for developers wanted to reuse an oracle, specially for those coming from classical software engineering, and therefore should be clearly documented as such. To make these functions more reusable, the user must clearly understand their interface (i.e., those properties of the function that interact with other artifacts) \cite{krueger1992software}.

To improve reuse these parameters should be clearly distinguished from other types of input parameters like those that are only used by the oracle and not by the function that creates it or those parameters used by both. An example of the first type is the specific qubits to which the oracle is going to be applied. This information is relevant for the oracle itself, and therefore it would be discussed in the next section, but is irrelevant for the function that creates the oracle. The quantum circuit for the range $[4, 7]$ would be the same regardless of whether it is applied to qubits from $q_0$ to $q_4$ or from $q_5$ to $q_8$.



\subsection{Reuse of the Oracle}

Finally, the last level of reuse to consider when creating oracles is the reuse of the oracle itself. Similar to classical software, using a formalized process to foster reuse, as the one proposed by these guidelines, increases the chance that the software can be reused successfully \cite{rothenberger2003strategies}. In the specific case of quantum oracles, some additional aspects should be considered that are not included in the reuse of classical software.

As part of classical software documentation, pre and postconditions are fundamentals aspects of software reuse. For oracles, one of the most relevant preconditions is the quantum state that the oracle is expecting as input. This is a key difference between implementations A and B. Whereas implementation A marks the states regardless of their amplitude, implementation B requires a superposition of all possible states with no relative phase. If the input does not meet this precondition, the oracle B will not behave as expected. In other oracles, the expected state could be different. 

As important as the input quantum state are the postconditions that can be guaranteed after the oracle has been applied to the quantum state. Specifically, the quantum state in which the qubits are left once the oracle has been applied. This can be extrapolated from the oracle's unitary matrix (as shown in Section \ref{sec:RangeOracles} for the range of integers example), however a textual description of the state would improve readability and foster reuse. For both implementations of the range of integer oracles, the postcontition of the quantum state is that states that represent integers in the selected range would have $\pi$-phase without any other change to the input quantum state (as long as input state fulfills the preconditions). Another example would be the addition oracle. The postcondition of this oracle is just a displacement on any given input state, i.e, for an input state $|4\rangle = |100\rangle$ an addition of 3 results in the state $|7\rangle = |111\rangle$. It also maintains the phases of the displaced states.


Therefore, an important guideline for creating reusable quantum oracles is to clearly document the pre and postconditions of the quantum state manipulated by the oracle. Providing this information will make oracles easier to reuse, since developers understand the expected state and how to keep working with the output state once the oracle has been applied.

Finally, the last proposed guideline for creating reusable quantum oracles is to document the properties of the oracles' quantum circuits that are relevant for future users.


One of the most relevant properties of the circuit to be documented, given the current state of quantum hardware, is the depth of the oracle's circuit. As has been discussed above, this property is crucial to determine if a given oracle can be successfully used in a NISQ era quantum computer and, therefore, it should be known by developers willing to reuse the oracle. The depth of the circuits should be measured by transpiling them to a specific device, as the architecture of quantum chips is not yet standardized and can greatly affect the resulting depth.

Nevertheless, other aspects of the circuit should also be considered. For example, the quantum gates used in the circuit are also a relevant aspect to consider if it is going to be run a real device. Although not available quantum gates can be replaced by alternative circuits, this could heavily impact the circuit's depth or other properties, so it should be documented for reuse. Something similar happens with the assumption of connections between given qubits for entanglement. Although this is usually handled by the transpiler it can affect the circuit properties. In general, any aspect of the circuit that could affect to its performance should be thoroughly documented. 

For the range of integer oracles, both implementations need the universal set of quantum gates. Moreover, all qubits require connections between them, because the less-than oracle and the Quantum Fourier Transform.

\subsection{Documentation of the range of integer oracles following the proposed guidelines}
\begin{figure*}[h]
    \centering
    \includegraphics[width=0.95\linewidth]{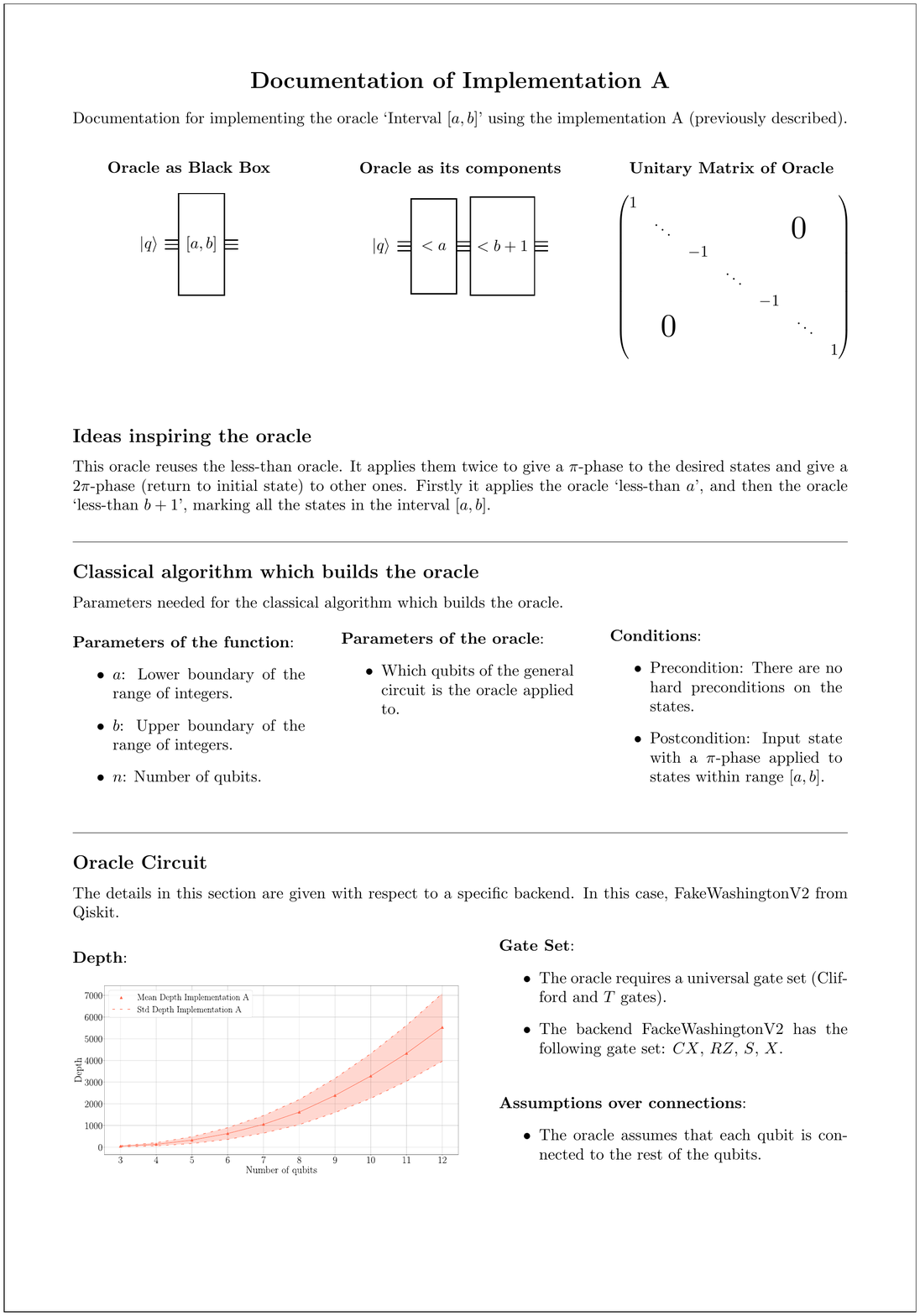}
    \label{fig:oracledocA}
\end{figure*}

\begin{figure*}[h]
    \centering
    \includegraphics[width=0.95\linewidth]{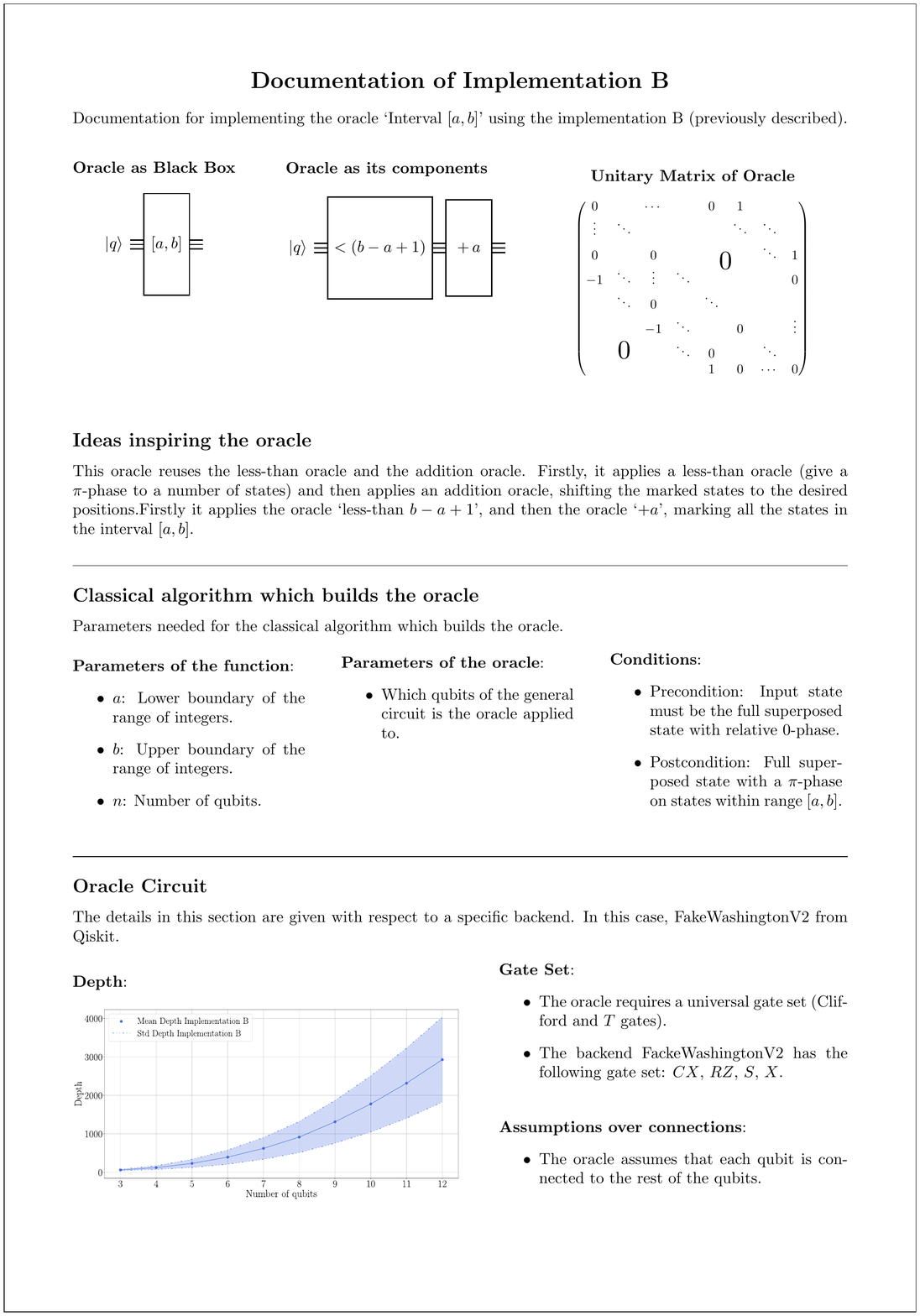}
    \label{fig:oracledocB}
\end{figure*}

The documentation cards of the two implementations can be found in \hyperref[fig:oracledocA]{documentation of implementation A} and \hyperref[fig:oracledocB]{documentation of implementation B}.

\section{Conclusions and Future Work}\label{sec:conclusions}
We have presented two different implementations of a range of integers oracle built by reusing two different simpler oracles, less-than oracle and addition oracle. This is done to exemplify the reusability of oracles. The functionality of both implementations has been shown with the same example. A study on the depth of both implementations has been made. An improvement in depth is clearly shown with one of the implementations. Moreover, we have presented several guidelines for making reusable oracles taking the range of integers oracle as an example, including the ideas inspiring the oracle, the function which creates the oracle and the oracle itself. We have provided an example of a documentation that follows those guidelines for both implementations presented.

This work aims to be a first point of discussion towards quantum software reusability. There is still work to do in establishing proper criteria for quantum software reusability, starting from the fact that this software is in an early development stage. Further guidelines may be presented not only on oracles, but also on any type of quantum software, such as full algorithm implementations.

\section*{Acknowledgments}
This work has been financially supported by the Ministry of Economic Affairs and Digital Transformation of the Spanish Government through the QUANTUM ENIA project call - Quantum Spain project, by the Spanish Ministry of Science and Innovation under project PID2021-124054OB-C31, by the Regional Ministry of Economy, Science and Digital Agenda, and the Department of Economy and Infrastructure of the Government of Extremadura under project GR21133, and by the European Union through the Recovery, Transformation and Resilience Plan - NextGenerationEU within the framework of the Digital Spain 2026 Agenda.

We are grateful to COMPUTAEX Foundation for allowing us to use the supercomputing facilities (LUSITANIA II) for calculations.

\section*{ Code and data}\label{sec:code-data}
The data used for generating the plots and the code for automatic generation of the oracles can be found in the following repository: \href{https://anonymous.4open.science/r/range-integers-oracle-C81E}{https://anonymous.4open.science/r/range-integers-oracle-C81E}.

\bibliographystyle{IEEEtran}
\bibliography{IEEEabrv,bibliography}

\end{document}